\documentclass[a4paper,11pt]{article}
\pdfoutput=1
\usepackage{jheppub}
\usepackage{textcase}
\usepackage{graphicx,epsfig,psfrag,amssymb,hyperref}
\usepackage{multirow}
\usepackage{color,graphicx,epsfig,psfrag,amsmath,empheq}
\usepackage[dvipsnames]{xcolor}
\usepackage{bm}
\usepackage{mathrsfs,amsfonts,soul,color}
\usepackage[caption=false]{subfig}
\usepackage{hepunits}
\usepackage{enumitem}
\usepackage{placeins}
\usepackage{textcomp}
\usepackage{tcolorbox}
\usepackage{float}
\usepackage[toc,page]{appendix}
\usepackage[ruled,vlined]{algorithm2e}
\usepackage{wrapfig}

\newcommand\snowmass{\begin{center}\rule[-0.2in]{\hsize}{0.01in}\\\rule{\hsize}{0.01in}\\
\vskip 0.1in Submitted to the  Proceedings of the US Community Study\\ 
on the Future of Particle Physics (Snowmass 2021)\\ 
\rule{\hsize}{0.01in}\\\rule[+0.2in]{\hsize}{0.01in} \end{center}}

\begin{document}

\title{Software and Computing for Small HEP Experiments\footnote{Authors have listed their experimental affiliations to illustrate the diversity of experiments represented here, but the opinions, findings, and conclusion or recommendations expressed in the paper are those of the authors and do not necessarily reflect the view of their collaborations.}} 

\author[1,a,b]{Dave Casper (editor),}
\author[2,3,4,c,d]{Maria Elena Monzani (editor),}
\author[5,b,e]{Benjamin Nachman (editor),}
\author[6,7,f,g]{Costas Andreopoulos,}
\author[5,h]{Stephen Bailey,}
\author[5,8]{Deborah Bard,}
\author[5,8]{Wahid Bhimji,}
\author[9,i]{Giuseppe Cerati,}
\author[10]{Grigorios Chachamis,}
\author[11,j]{Jacob Daughhetee,}
\author[12,k]{Miriam Diamond,}
\author[9,l]{V. Daniel Elvira,}
\author[2,3,c]{Alden Fan,}
\author[9,m,n,o]{Krzysztof Genser,}
\author[13,p]{Paolo Girotti,}
\author[5,c]{Scott Kravitz,}
\author[9,m]{Robert Kutschke,}
\author[14]{Vincent R. Pascuzzi,}
\author[9,q]{Gabriel N. Perdue,}
\author[9,i]{Erica Snider,}
\author[9,n]{Elizabeth Sexton-Kennedy,}
\author[15,b,n]{Graeme Andrew Stewart,}
\author[16,c]{Matthew Szydagis,}
\author[17,a,b]{Eric Torrence,}
\author[18,r]{Christopher Tunnell}


\affiliation[1]{University of California, Irvine}
\affiliation[2]{SLAC National Accelerator Laboratory}
\affiliation[3]{Kavli Institute for Paritcle Astrophysics and Cosmology}
\affiliation[4]{Vatican Observatory}
\affiliation[5]{Lawrence Berkeley National Laboratory}
\affiliation[6]{University of Liverpool}
\affiliation[7]{STFC Rutherford Appleton Laboratory}
\affiliation[8]{National Energy Research Scientific Computing Center (NERSC)}
\affiliation[9]{Fermi National Accelerator Laboratory}
\affiliation[10]{Laborat{\' o}rio de Instrumenta\c{c}{\~ a}o e F{\' \i}sica Experimental de Part{\' \i}culas (LIP)}
\affiliation[11]{Oakridge National Laboratory}
\affiliation[12]{University of Toronto}
\affiliation[13]{INFN Pisa}
\affiliation[14]{Brookhaven National Laboratory}
\affiliation[15]{CERN}
\affiliation[16]{SUNY Albany}
\affiliation[17]{University of Oregon}
\affiliation[18]{Rice University\vspace{3mm}}

\affiliation[a]{FASER Collaboration}
\affiliation[b]{ATLAS Collaboration}
\affiliation[c]{LZ Collaboration}
\affiliation[d]{Fermi-LAT Collaboration}
\affiliation[e]{H1 Collaboration}
\affiliation[f]{T2K Collaboration}
\affiliation[g]{SBND Collaboration}
\affiliation[h]{DESI Collaboration}
\affiliation[i]{MicroBooNE Collaboration}
\affiliation[j]{COHERENT Collaboration}
\affiliation[k]{SuperCDMS Collaboration}
\affiliation[l]{CMS Collaboration}
\affiliation[m]{Mu2e Collaboration}
\affiliation[n]{HEP Software Foundation}
\affiliation[o]{Geant4 Collaboration}
\affiliation[p]{Muon g-2 Collaboration}
\affiliation[q]{MINERvA Collaboration}
\affiliation[r]{Xenon Collaboration}

\emailAdd{dcasper@uci.edu}
\emailAdd{monzani@stanford.edu}
\emailAdd{bpnachman@lbl.gov}

\abstract{
This white paper briefly summarized key conclusions of the recent US Community Study on the Future of Particle Physics (Snowmass 2021) workshop on Software and Computing for Small High Energy Physics Experiments.
\snowmass
}

\maketitle

\section{Introduction}
\label{sec:intro}

This white paper summarizes the discussion from a dedicated Snowmass workshop on Software and Computing for Small High Energy Physics (HEP) Experiments (\url{https://indico.physics.lbl.gov/event/1756}).  The mandate for the workshop was:

\begin{itemize}
\item Identify unique computational challenges of the `small' experiment community.
\item Gather input about what is needed in terms of computation for these experiments to be successful.
\item Connect members of the `small' experiment community to the computational frontier in Snowmass and encourage participation in topical groups.
\item Foster the development and re-use of open-source software, building on the work of the HEP Software Foundation (HSF) and other collaborative efforts within the community.
\end{itemize}
In order to be inclusive, we did not impose a definition of `small' and have asked experiments to self-select.  The workshop was organized into two days: a first day for perspectives from representative experiments and a second day about computational tools.  In particular, on the first day, we heard from CERN-based experiment FASER~\cite{Feng:2017uoz}, Fermilab-based experiments MicroBooNE~\cite{MicroBooNE:2016pwy} and g-2~\cite{Muong-2:2015xgu}, and non-CERN/FNAL experiments LUX-ZEPLIN (LZ)~\cite{LZ:2019sgr}, COHERENT~\cite{COHERENT:2018gft}, and DESI~\cite{DESI:2019jxc}.  These experiments covered all of the HEP frontiers.  On the second day, there were speakers covering the HSF, High Performance Computing (HPC), event processing frameworks, reconstruction frameworks, event generators, detector simulations, and machine learning.  The slides contain a wealth of information about each of these topics and are preserved at the indico link above.  The goal of this white paper is to describe some of the key takeaway points from the presentations and discussions.  If the entire workshop had to be summarized in one phrase, it would be\footnote{Quoted from E. Snider's presentation.}

\begin{tcolorbox}[colback=white]
\begin{align}\nonumber
\textbf{small experiment} \neq \textbf{small data volume or small computing problem!}
\end{align} \vskip 1cm
\end{tcolorbox}

\section{Key Observations and Recommendations}
\label{sec:recs}

Below is a summary of important points discussed at the workshop; key recommendations are \textbf{offset in bold}.  The four subsections below are in no particular order.

\subsection{High Performance Computing Centers}

Increasingly, small experiments need to rely on HPC resources for a significant fraction of their computing budget.  At the workshop, we heard about challenges by experiments in porting to HPC resources and about challenges from HPC center staff about accommodating small experiment users.

From the perspective of an HPC center,  `small' experiments look very similar to `large' experiments. 
They have similar scales in terms of compute hours, storage used and number of users, and they all need peripheral services such as databases and workflow controls, as well as allocations on the big systems. 
For an HPC center, this means both `small' and `large' experiments present the same demands on resources. 
What distinguishes `small' experiments is that many of the common issues faced by all experiment teams working at HPC centers are exacerbated by the lack of human resources available to work on these issues. 
It is vital that science teams and HPC centers partner to address these challenges. 
These issues fall into roughly three areas:  
\begin{itemize}
    \item Getting codes running on new architectures
    \item Adapting to HPC center policy
    \item Tools that are easy to stand up and maintain
\end{itemize}

\subsubsection{Key challenge: Porting workflows to new architectures}
The DOE Office of Advanced Scientific Computing Research (ASCR) operates three computing and one networking user facilities. 
All have a joint mission to advance both science and the state of the art in computing. 
ASCR facilities engage with vendors years in advance of a system's delivery, to ensure that production systems push the limits of existing technology. 
There are three key drivers in this advancement that pose a challenge to scientific workflow design: 
\begin{itemize}
    \item Energy efficiency. Scientific computing needs are ever-increasing, but there is a limit to the amount of power and cooling we can safely bring into a building. Energy efficient architectures are here to stay, including specialized devices that perform specific calculations at ultra-high efficiency (e.g., for AI/ML).
    \item New storage technologies. Object stores are increasingly prevalent, and massively parallel file systems do not always respond well to data-intensive workloads.
    \item Peripheral services. We are seeing an increasing use and reliance on containers, scalable databases, and interfaces to other computing sites (e.g., local clusters, the cloud).
\end{itemize}

A pressing question for both experiment teams and HPC center staff is how to optimize the entire end-to-end workflow, not just individual applications. 
The NESAP program at NERSC (the National Energy Science Research Computing center) partners with application development teams and vendors to port and optimize codes to new architectures and platforms. 
Lessons learned from this process are shared with the wider NERSC community via documentation and training. 
NESAP is evolving with the types of systems we deploy, and will need to support workflow optimisation in the future. 

\textbf{Programs like NESAP are extremely valuable to small experiments and we advocate for a continuation and/or expansion of this program}.  However, we also heard about difficulties in recruiting NESAP postdocs.  This is discussed in more detail in Sec.~\ref{sec:workforce}.

\subsubsection{Key challenge: Policy optimization}
Some of the biggest challenges faced by users of HPC centers are due to security and policy concerns, rather than technical barriers. 
Supporting the needs of experiment workflows has pushed NERSC to change policies around (for example) near-realtime access to systems, and support for federated ID. 

In addition, NERSC has recognised that some workflows highly value continued access to NERSC services - for example, if an experiment is operating 24/7. 
Through careful planning and investment, we are now able to keep most infrastructure up during outages using generators and new system software capabilities for rolling upgrades. 
We are also supporting research in several projects to support more portable cross-facility workflows~\cite{alcc}. 

\subsubsection{Key challenge: Scaling user support}
NERSC is increasingly supporting more users and projects from experiment facilities (including other DOE user facilities). 
HPC centers are facing a sea change in the number and type of users they support - without any change in the number of staff available to support them. 
This challenge requires a new approach to user support, to make it easier for users to develop and adapt their workflows without direct support from NERSC staff. 
If done successfully, this will also make it easier for `small' experiments to adapt to HPC centers. 

The LBNL Superfacility Project~\cite{superfacility} was designed to leverage and integrate work being done across NERSC, ESnet and research divisions at LBNL to provide a coordinated and coherent approach to supporting experimental science at DOE facilities. 
The principles behind the project were to provide an integrated, scalable and sustainable framework for experiment science, working closely with a range of science teams to get the design right and using existing, industry standard and open source tools wherever possible. 
This 3-year Project was completed at the end of 2021, having achieved its aim, although work continues very actively on Superfacility topics. 
NERSC is now able to support automated pipelines that analyze data from remote facilities at large scale, without routine human intervention, using capabilities such as near-realtime computing, dynamic networking, API-driven automation, HPC-scale notebooks with Jupyter~\cite{jupyter}, state of the art data management tools, federated ID and container-based peripheral services.

\subsection{Requirements beyond FLOPS, bytes, and bandwidth}

Classic high performance computing (HPC) centers have traditionally focused on maximizing the number of
floating point operations per second (FLOPS) delivered to massively
parallel programs, coupled with large bandwidth to disk and tape storage.
Small experiments often have requirements that go beyond just
FLOPS, bytes, and bandwidth, e.g.~databases, data access portals,
I/O robustness, complex workflows with many interdependent smaller jobs,
collaboration accounts accessible by multiple users to coordinate data processing, queue waittime, and jupyter notebook servers.
High uptime, low mean time to failure (MTTF), and access to realtime and interactive computing resources are particularly important for the effective use of computing centers by smaller experiments.
Supporting these requirements takes tangible specific effort from the computing centers, and tangible specific effort from the experiments to adapt.
As such, HPC centers need dedicated support for non-HPC workloads with
appropriate metrics for evaluating success, e.g. minimizing MTTF and providing interactive access, and not just FLOPS delivered.
Similarly, experiments need dedicated support to adapt and maintain their workflows to take full advantage of the opportunities at HPC centers.

\subsection{Software Challenges}
\begin{figure}
    \centering
    \includegraphics[width=1.0\textwidth]{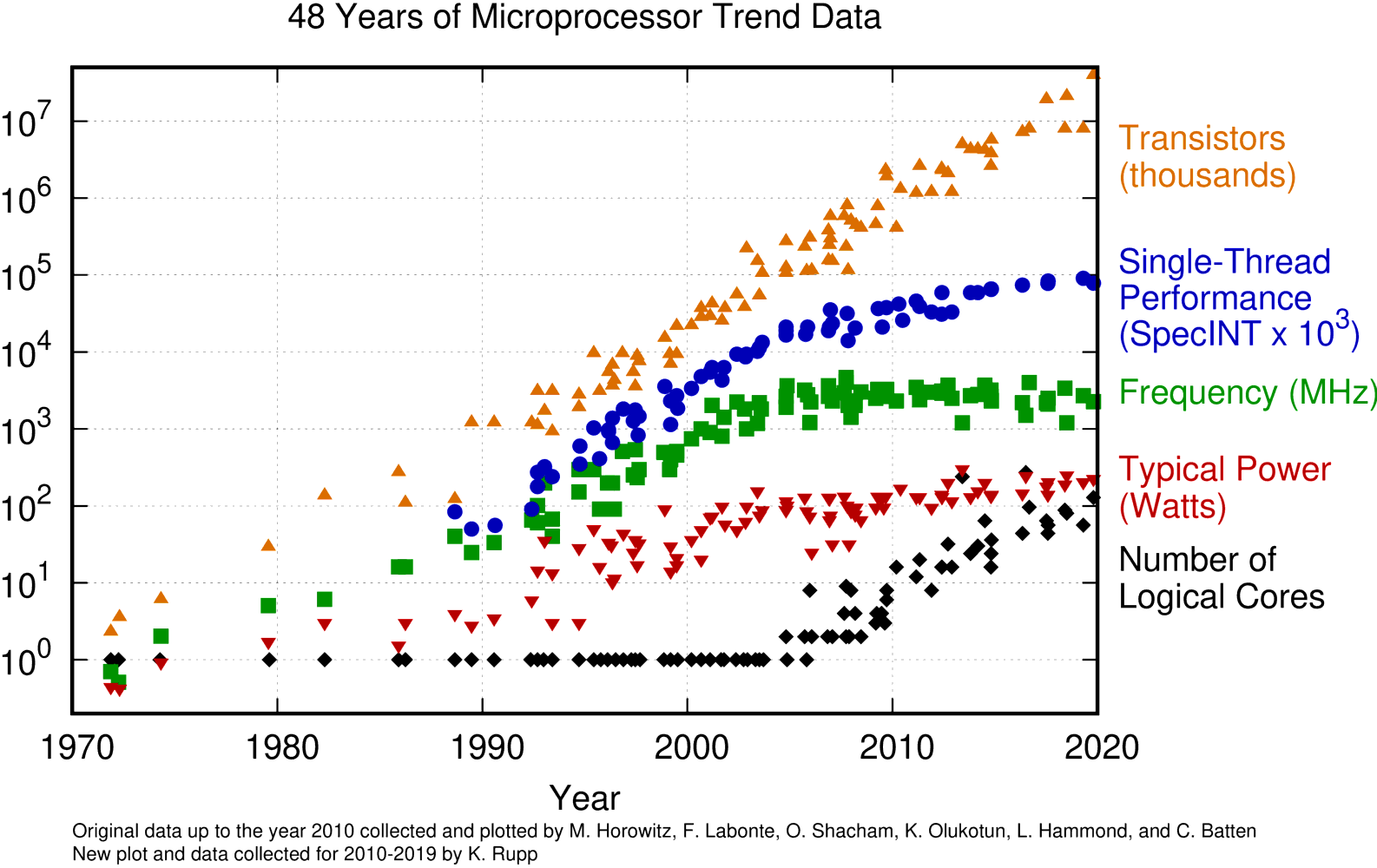}
    \caption{Microprocessors reach the power wall limit circa 2010}
    \label{fig:Microprocessor_trend}
\end{figure}

High Energy Physics has a vast investment in software estimated to be greater than 50M lines of C++ code. It is a critical part of our physics production pipeline, from triggering and simulation all the way to analysis and final plots. This legacy represents a huge investment over the years.  It requires ongoing support in human effort to adapt to changing computing technologies. As shown in figure ~\ref{fig:Microprocessor_trend}, improved single-threaded performance will not come from the hardware.  Computing gains come from increasing concurrency in the code, especially at HPC centers. Concurrent programming skills must be learned
and not all experimenters want to invest in becoming proficient in concurrent programming.  There is no longer a large supply of graduate students and postdocs, the historical sources of experiment software, that can write or even maintain codes for these complicated architectures. This hits smaller experiments especially hard, because there may not be a critical mass of people with the critical skills needed. See Sec.~\ref{sec:workforce} for suggestions on how to address personnel needs.
Efforts to share code developed by the large experiments should be funded to have broader impact for the community, and thereby maximize return on those investments.   This means adding time for documentation, training, and outreach. Whenever possible the code should be open sourced. 

\subsection{Support for Common Tools}

A reoccurring theme in nearly all of the experiment vignettes was the need for common tools.  A distinguishing feature of small experiments compared to large ones is that they often do not have many, many or, in some cases, any dedicated software and computing experts. This means that there is limited/no resources to develop new data acquisition, simulation and reconstruction tools from scratch as large experiments are often able to do. Some small experiments benefit from shared expertise with large experiments, especially those located at CERN and FNAL. Yet many small experiments do not have such natural overlap, and they are left with few avenues to obtain the resources necessary to develop their required suite of tools.  

In some cases, larger experiments have (at least partially) abstracted and open-sourced their software in a way that it can be re-used by small experiments. Two great examples of this are the software packages ACTS~\cite{Ai:2021ghi} and LArSoft~\cite{Snider:2017wjd} (\url{https://larsoft.org}) as well as the event processing frameworks Gaudi~\cite{Barrand:2001ny}, Art~\cite{Green:2012gv} and Stitched (\url{https://github.com/cms-sw/Stitched}), an abstraction of the CMSSW framework\cite{Jones:2006,Jones:2014,Jones:2015,Jones:2017,Bocci:2020}).  
ACTS (A Common Tracking Software) is a package for tracking at collider experiments, initially based on ATLAS code. It provides experiment-independent implementations of track and vertex reconstruction algorithms which are available for usage by any experiment once their geometry and material description is implemented in ACTS. Successful examples include ATLAS, FASER, sPHENIX, Belle-II. LArSoft is a shared base of simulation, reconstruction, and analysis software across liquid argon time projection chamber (LArTPC) neutrino experiments. LArSoft is currently used by all Fermilab-based LArTPC experiments, including DUNE and smaller-scale experiments (SBN, MicroBooNE, LArIAT, ArgoNeuT). Both ACTS and LArSoft provide ready-to-use software solutions as well as a platform for R\&D work spanning usage of AI methods, algorithm parallelization, and deployment at HPC, which can significantly reduce the required workload for smaller experiments.

However, some open source tools still require significant adaptation or development and thus require experiment-independent support.  The best example of this is the widely-used detector simulation software Geant4~\cite{GEANT4:2002zbu,Allison:2006ve,Allison:2016lfl}.  Even for large experiments, significant resources are required to integrate Geant4 into the detector-specific setup.  Many small experiments have physics needs beyond those covered by large experiments and additional development is necessary. Some of this is happening organically (e.g. NEST~\cite{nest} for direct dark matter detection), but this is not always the case. To strengthen the partnership between the Geant4 Collaboration and the experiments, Geant4 has recently introduced a  ``Geant4 contributor'' status, which should facilitate a faster inclusion of codes developed by the experiments, benefiting the entire community. 

In the US, the HEP support for Geant4 comes from the computing operations budgets of large experiments and intensity frontier operations at Fermilab. This leads to a stove-piping of Geant4 support such that issues specific to those experiments are addressed because they are needed for the operation of the experiment. Other experiments that use Geant4 are left out in the cold. In addition, Geant4's common software, such as physics models, no longer receives any US maintenance funding. The physics Generators used in the field (eg. Pythia, GENIE, Madgraph, Sherpa) also suffer from lack of stable funding in a similar way. This is not sustainable for the long term viability of small experiments. \textbf{Long term, experiment-agnostic Geant4 support is critical for the success of small experiments.} At the workshop, we also heard that many small experiments do not update Geant4 versions frequently (or ever!), and so they are unable to take advantage of new developments, whereas maintaining many old versions by the Geant4 collaboration is unsustainable. This means that small experiments often use no longer supported Geant4 versions and miss out on recent physics developments and computational innovations, such as multi-threading or the ongoing integration with Graphical Processing Units (GPUs).

\subsection{Data Storage and Preservation}

Data preservation is necessary for scientific repeatability, while helping maximize the ``return on investment'' of our experiments.  Re-analysis of data is of particular interest, as advances in analysis methods and physics models become available such as more detailed simulations and new applications of machine learning. Well-understood datasets are often the best source of training for young physicists, and open-access datasets support outreach efforts by targeting both data literacy and science skills. In addition, standardized, trusted data repositories are critical for developing and testing new machine-learning algorithms~\cite{Dua:2019}. Encouraging the experiments to make their data publicly available will lead to a richer ecosystem of available datasets, and help with long-term preservation of data and experimental knowledge. Most important is the publication of data following the end of an experiment.  One way of ensuring data are usable for post-experiment use is to periodically release data throughout the lifetime of the experiment.  Another model is to internally future-proof data formats and analysis software to always be able to analyze older datasets.  In both cases, person-power is required for an effective implementation. 

Computing clusters do not typically guarantee hosting for significant lengths of time, nor do they have mechanisms to easily share these data with the public. However, services for long-term data and software preservation (\url{https://heasarc.gsfc.nasa.gov} and \url{https://lambda.gsfc.nasa.gov}) have been used for decades in astronomy, and the HEP community has become quite familiar with them. Ensuring data usability long after the end of an experiment is not quite as simple as ``dumping everything'' in a publicly available repository. Detailed detector information is needed, as well as access to metadata such as slow-control and calibration databases. Moreover, community standards specifying data format and interfaces must be agreed upon. Finally, prototype analysis codes must be published with the datasets, further lowering the barrier to data access and ensuring that the data is analyzed correctly. It has become more commonplace in our community to take advantage of general purpose data and code services like Zenodo (\url{https://www.zenodo.org}), which provide most of the infrastructure required for effective data preservation.

\subsection{Computing Personnel}
\label{sec:workforce}

The general challenge of maintaining software and computing talent is exacerbated in small experiments by the lack of long term, permanent positions within the experiments. Many of the software tasks are carried about by young collaborators and there is significant turnover from year to year. A special case is an experiment that is both small and far from data, for which it is common that the community of junior people working on software is not large enough to be self-sustaining. This problem can be partly mitigated by increased use of common tools but that does not address the task of developing experiment specific configurations, customizations and glueware.

It is therefore essential that there is \textbf{funding for permanent software and computing experts}. The careers of these researchers should be evaluated appropriately with rewards to efficiency, stability, and robustness. A more transparent approach to software development and data sharing would go a long way towards improving the career prospects of software and computing experts, as it would allow individuals to claim credit for their work and be evaluated appropriately. There is little funding for software maintenance, even though this is critical. One possibility is the creation of research software engineer (RSE) positions that have long term funding independent of experiments, but for the support of existing and planned experiments. Another area of opportunity is the increase of joint particle physics and data science appointments at universities, which have become marginally more common over the last decade. We also need to increase the community investment in general software literacy of physicists, through (continuing) education initiatives and collaborations with industry, national labs and academia.

The diversity issues pervasive in HEP are exacerbated in the computing domain due to the (perceived) technical nature of the work, and we must ensure that faculty, staff, and trainee positions are viable career options for a diverse group of people. Our workforce development efforts should explicitly include equity, and recognize that diversity is foundational to our success, as it demonstrably increases the creativity of solutions and variety of approaches. Diversifying the computational workforce will require efforts to diversify physics in general, and to leverage our partnership with the astronomy and industry communities, who have made substantial efforts in this direction over the last decade. This remains a challenging problem, but not insurmountable, as STEM identity formation for underrepresented minorities has become a priority in a variety of educational settings, from early childhood to higher education~\cite{STEM-diversity}.

\section*{\label{sec::acknowledgments}Acknowledgments}

We thank the participants of the Software and Computing for Small HEP Experiments workshop \url{https://indico.physics.lbl.gov/event/1756} for many useful discussions.

This work was supported by the US Department of Energy, Office of Science, under contract numbers DE-AC02-76SF00515 (SLAC) and DE-AC02-05CH11231 (LBNL), and by the Fermi National Accelerator Laboratory, managed and operated by Fermi Research Alliance, LLC under Contract DE-AC02-07CH11359 with the US Department of Energy. GC acknowledges support by the Funda\c{c}{\~ a}o para a Ci{\^ e}ncia e a Tecnologia (Portugal) under project CERN/FIS-PAR/0024/2019 and contract 'Investigador auxiliar FCT - Individual Call/03216/2017' and from the European Union's Horizon 2020 research and innovation programme under grant agreement No. 824093.

\bibliographystyle{JHEP}
\bibliography{main,HEPML}

\end{document}